\documentstyle[11pt,newpasp,twoside]{article}
\newcommand{\ergs}{\rm \ erg \; s^{-1}}

\def\edcomment#1{\iffalse\marginpar{\raggedright\sl#1\/}\else\relax\fi}
\reversemarginpar

\begin{document}
\title{Variability in the quiescent state of neutron star transients}
\author{Sergio Campana}
\affil{INAF-Osservatorio astronomico di Brera}

\begin{abstract}
Variability studies represent a powerful tool to assess the emission
mechanism(s) at work in the quiescent state of transient low mass X--ray binaries. 
Here we report on Chandra and XMM-Newton observations of two well known
transient sources: Aql X-1 and Cen X-4. Long and short term
variability is observed. This variability can be easily interpreted in view of
an active millisecond radio pulsar emitting X--rays at the shock between a
radio pulsar wind and inflowing matter from the companion star.
\end{abstract}

\section{Introduction}

Many Low Mass X--ray Binaries (LMXRBs) accrete matter at very
high rates, and therefore shine as bright X--ray sources, only sporadically.
Among these systems are Soft X--ray Transients (SXRTs) hosting an
old neutron star (for a review see Campana et al. 1998).
These systems alternate periods (weeks to months) of high X--ray luminosity,
during which they share the same properties of persistent LMXRBs,
to long (1--few 10 years) intervals of quiescence in which the X--ray
luminosity drops by up to 5--6 orders of magnitude.

Short term variability is a powerful tool for the study of the emission
mechanism(s) responsible for the SXRTs quiescent emission.
A factor of 3 variability over timescales of days (Campana et al. 1997)
and of $40\%$ over 4.5 yr (Rutledge et al. 2001) has been reported in Cen X-4.
Several other transient neutron star systems have also been found to be
variable in quiescence by factors of 3--5 (e.g. Rutledge et al. 2000) but data
have been collected over several years and with different instruments.
Here we report on recent results obtained on two of the best studied sources
of this class.

\section{Aql X-1}

Rutledge et al. (2002) analysing Chandra data of the Aql X-1 quiescent phase
after the November 2000 outburst found a variable flux and X--ray spectrum.
They interpreted these variations in terms of variations of the neutron star
effective temperature, which changed from $130^{+3}_{-5}$ eV, down to
$113^{+3}_{-4}$ eV, and finally increased to $118^{+9}_{-4}$ eV, suggesting
low level accretion to continue also in quiescence.

Chandra data on Aql X-1 are the first that show a clear luminosity variation
and, more importantly, an increase of the temperature during quiescence. 
Campana \& Stella (2003)  approached the same Chandra data plus an unpublished
long BeppoSAX observation carried out in the same
period to probe a different spectral model.
Deep crustal heating (Brown et al. 1998; Colpi et al. 2001) has been proposed
as a physically sound 
mechanism powering the soft component of the quiescent spectra of SXRTs.
Several mechanisms have been proposed to explain the hard tail component often
observed in quiescent SXRTs. One of them, physically motivated by the recent
observations of the millisecond radio pulsar (MSP) PSR J1740--5340 (D'Amico et
al. 2001; Grindlay et al. 2001), relies on the shock emission between the
relativistic MSP wind and matter outflowing from the companion (Tavani \&
Arons 1997; Campana et al. 1998). These two components are not exclusive.

Within this physical scenario, we fit the spectra fixing the soft component for
all the observations and leave free to vary the hard component {\it and} the
column density (that changes by a factor of $\sim 5$).
This model is consistent with the
entire dataset ($\chi^2_{\rm red}=1.00$, for 109 d.o.f. and with a null
hypothesis probability of $49.0\%$). Fitting the same dataset with the
best fit model by Rutledge et al. (2002) with the addition of the BeppoSAX
data, we obtain a slightly worse fit ($\chi^2_{\rm red}=1.17$, for 113
d.o.f. and with a null hypothesis probability of $11.1\%$).
We conclude that the scenario proposed is (at least) equally
well consistent with the data, meaning that a shock emission scenario can
account for the spectral variability observed in Aql X-1.
We also note that Rutledge et al. (2002) found $32\%$ (rms) variability in
last Chandra observation. In their case the power law component contributed
only $12\%$ of the flux. From our fit the power law component contributes to
$38\%$ of the total flux, so it can in principle account for all of the
short-term variability.

Despite the low number of points, spectral parameters derived for the power
law index show some correlation with the column density (interpreted as a
measure of the variable mass around the system, over a fixed interstellar
amount) as well as with the power law flux. This correlation might be expected in
the shock emission scenario (Tavani \& Arons 1997).
What is now expected is the large value of the power law index in the last
observations. This might then provide an indication of a different regime in
the system, possibly underlying a larger inverse Compton cooling. 

\section{Cen X-4}

Thanks to XMM-Newton large throughput, Cen X-4 was observed at the
highest signal to noise ever. This allowed us to disclose rapid ($>100$ s),
large ($45\pm7\%$ rms in the $10^{-4}-1$ Hz range) intensity variability,
especially at low energies (Campana et al. 2004). 
The power spectrum can be well fit with a power law with
index $-1.2\pm0.1$. No significant periodicities or quasi-periodic
oscillations are seen. The power-law index of the X--ray power spectrum
is consistent with the optical one which is characterized by an index of $\sim
-1$, $-1.5$ (Hynes et al. 2002; Zurita et al. 2003).

In the pn light curve three flare-like events can be identified.
Flare activity has been reported also in the optical for a number of
transient black holes during quiescence as well as for Cen X-4 (Zurita
et al. 2003; Hynes et al. 2002). In the optical flares occur on timescales of
minutes to a few hours, with no dependence on orbital phase. R-band
luminosities are in the range $10^{31}-10^{33}\ergs$. The mean duration of
optical flares in Cen X-4 is 21 min. This is similar to what observed in the
X--rays. 

In order to highlight the cause of this variability, we divided the data into 
intensity intervals and fit the resulting spectra with the
canonical model for neutron star transients in quiescence, i.e. an absorbed
power law plus a neutron star atmosphere. 
Small spectral variations are observed.
These can be mainly accounted for by a variation in the column density together
with another spectral parameter. Based on the available spectra we cannot
prefer a variation of the power law versus a variation in the temperature of
the atmosphere component (even if the first is slightly better in terms of
reduced $\chi^2$).

\section{Conclusions}

Variations in the temperature of the neutron star atmosphere might suggest
that accretion onto the neutron star surface is occuring in quiescence;
variations in the power law tail (photon index) together with variations in
the column density should support the view of an active millisecond radio
pulsar emitting X--rays at the shock between a radio pulsar wind and inflowing
matter from the companion star. We tested both models with state of the art
data on Aql X-1 (long term variability with Chandra) and Cen X-4 (short term 
variability with XMM-Newton). Despite current toughts either model can explain
equally well the observational data.

\end{document}